\documentclass[preprint,pdf]{iucr}              
\usepackage{graphicx}
\usepackage{multirow}
\usepackage{float}
\usepackage{amsmath}
\usepackage{lineno}

     \journalcode{J}              

\begin{document}                  



\title{Reproducible x-ray reflectometry optimization: statistical analysis of differential evolution fitting of multilayer structural models}
\shorttitle{Reproducible XRR data optimization}


\cauthor[a]{Donald}{Windover}{windover@nist.gov}{address if different from \aff}
\author[b]{David}{Gil}
\author[c]{Yasushi}{Azuma}
\author[c]{Toshiyuki}{Fujimoto}

\aff[a]{National Institute of Standards and Technology, Gaithersburg, MD \country{USA}}
\aff[b]{Coruscavi Software, Washington, DC \country {USA}}
\aff[c]{National Institute of Advanced Industrial Science and Technology, National Metrology Institute of Japan, Tsukuba \country {Japan}}

\shortauthor{Windover et al.}







\maketitle                        

\begin{synopsis}
X-ray reflectometry modeling study for determining the amount of computer time needed to fit model to data. 
\end{synopsis}

\begin{abstract}
We test the reproducibility of X-ray reflectometry(XRR) measurements and optimizations using an National Metrology Institute of Japan (NMIJ)/National Institute of Advanced Industrial Science and Technology (AIST) pre-standard. 
Based on bootstrap analysis of repeated refinements, using several CPU-years of time, we provide concrete recommendations of best practices for ensuring the reproducibility
of XRR model fitting results. These recommendations can be used to study both instrument repeatability and cross-instrument reproducibility. Because the recommendations used optimizations methods available in
commonly used commercial software, they can quickly be applied both in research and analytical
laboratories, as well as fabrication environments.
\end{abstract}


\section{Introduction}

X-ray reflectometry (XRR) relies on the subtle differences in the index of refraction, $n_{layer}$, among layers in a multilayer stack to recover thickness and electronic density information from X-ray specular reflectivity patterns.  The formal derivation of the method is covered elsewhere (see \citeasnoun{lekner87}) and the XRR technique has been used extensively in the analysis of polymers \cite{russell90} and other materials systems \cite{chason97}. Although ubiquitously to study films in all materials disciplines, XRR is presently of vital interest for fabrication-line tool development to be used for process optimization in the semiconductor industry \cite{nolot12}.

XRR interference patterns are directly linked to layer thicknesses (see section \ref{section:parratt}) and provide more accurate thickness determination than other methods available today (such as visible-light ellipsometry, where thickness is coupled to pronounced changes in $n_{layer}$). \cite{archer62,irene93}  However, there is no guarantee that XRR will have high sensitivity for a given layer, as the method relies on high electron density contrast between layers \cite{ferrari04}.  The method also relies on both low roughness at interfaces, and uniform layer scattering densities; interdiffusion layers wreak havoc with method sensitivity \cite{windover05}.  The National Institute of Standards and Technology (NIST) has been working over the past decade to establish parameter uncertainty
\footnote[1]{Our goal is to estimate parameter uncertainties.  However, this work represents a first step of estimating precision of replicate measurements and inter-tool measurement reproducibility.  Combined uncertainty analysis will require a more comprehensive treatment of systematic effects introduced by measurement systems.\cite{vim2012}}
estimates for the XRR method, by determining robust model invariant parameters \cite{windover07} and by assessing the impact of one of the most common problems in XRR analysis, surface contamination \cite{gil12}.

The international community has performed several round-robins on XRR measurements through the Versailles Project on Advanced Materials and Standards (VAMAS) to establish inter-tool comparability \cite{colombi08,matyi08}.  In the first round-robin, a GaAs/AlAs bilayer, repeated 3 times, epitaxial structure, deposited on GaAs (and produced by the National Metrology Institute of Japan (NMIJ)/National Institute of Advanced Industrial Science and Technology (AIST) ) was used for the inter-comparison measurements.  This structure required a seven slab XRR model ( $2$ $\times$ $3$ $+$ $1$ surface contamination layer) introducing over 21 ($>7 \times 3$) possible free modeling parameters (assuming thickness ($t$), density($\rho$), and roughness ($Rz$) refinement per layer).  The intercomparison concluded that there was indeed high inter-tool uniformity and stability in thickness determination from the buried layers, however there was also a high variability in results for thickness using different modeling methods. 

NIST is working in collaboration with NMIJ/AIST to develop instrument  alignment procedures which will use certified reference materials (CRM)s to reduce inter-tool measurement uncertainties. In order to align an instrument using such an artifact, we must first establish a protocol for modeling XRR data, and determine baseline precision estimates for both the modeling method and for those caused by measurement noise.  Ultimately, these modeling dependent precision estimates must be separable from the reproducibility bias introduced by different measurement instrumentation, system configurations, and system alignments (or misalignments).  In this work, we perform modeling and data quality precision analysis for a structure very similar to the VAMAS material, which sheds some light on their findings. 
We will also take a first look at inter-tool reproducibility by analyzing XRR data from two instruments with nominally identical optical configurations. This modeling precision study is the first step towards establishing the efficacy of CRM structures in instrument alignment protocols.

\section{Discussion}

In order to establish a measurement and data analysis procedure, we will review all the tools used in performing the difficult task of evaluating precision estimates using only optimization methods:

\begin{itemize}
    \item How we measure the XRR data
    \item How we simulate XRR data for comparison  
    \item What fitness function we use
    \item What strategy we use in optimization
    \item How we turn optimizations into statistical results
\end{itemize}

\subsection{XRR measurement}
\label{section:measurement}
XRR measurements were made on two commercial Rigaku SmartLab\footnote[2]{Certain commercial equipment, instruments, or materials are identified in this paper in order to specify the experimental procedure adequately. Such identification is not intended to imply recommendation or endorsement by the National Institute of Standards and Technology, nor is it intended to imply that the materials or equipment identified are necessarily the best available for the purpose.} diffractometers, each using a graded parabolic multilayer optic and a Ge (220) 2-bounce monochromator, providing a parallel beam with high intensity (over $1\times10^7$ the detector background). 
Two instruments, one with a sealed tube (2.2 Kw), and one with a rotating anode (9 kW), were used.
XRR data are collected as a series of incident angle, $\theta^i$, and reflected intensity, $I^R$, data pairs, $(\theta^i_l, I_{l}^R)$ stepped over $N$ points in a range, $l=1, 2, \ldots, N$ from a starting incident angle, $\theta^i_{1}$ to ending incident angle $\theta^i_{N}$.   
Data was taken with $\theta^i_{1} = 0^{\circ}$, $\theta^i_{N} = 3.5^\circ$, $N = 1400$, and with count times of 1 s, 20 s, or 30 s per data point, to provide different count variance levels; single XRR scan times ranged from fast (22 minutes for the 1 s data set) to detailed (overnight for the 20 s and 30 s data sets).  
We present results from: ten XRR measurements at 1 s per point \& rotating anode instrument (1 s set $\equiv\{09,15,19,25,29,34,44,50,54,59\}$) [Scan labels are derived using data time stamp (2 digit starting minute for each)] , six XRR measurements at 30 s per point \& sealed tube instrument (30 s set  $\equiv\{22,23,26,43,58,00\}$), and three at 20 s per point \& rotating anode instrument (20 s set $\equiv\{26,40,43\}$).  
In all cases, the sample was aligned using a series of automated measurements (``precise sample alignment'' mode in SmartLab Guidance, version 1.5.5.3) to find the specular condition between detector, sample, and source axes ($\theta \cong \theta^i \cong \theta^R$). The instrument alignment uncertainty is on the order of $\delta \theta = \theta^i - \theta^R < 0.001^{\circ}$ (verified through preliminary repeatability studies).

The structure measured was a pre-standard (Lot \# BAAA4002C / 1-08) produced by NMIJ / AIST consisting of three bilayers of GaAs/AlAs, deposited using molecular beam epitaxy, on a single crystal GaAs wafer substrate. Each layer is roughly 9.5 nm in thickness.  Prior work has shown this structure has atomically smooth transitions between epitaxial layers, and that the stoichiometry and thickness of the buried layers is stable.  These pre-standards were produced in 2004, making them an ideal test structure for long-term stability studies. Over the decade since its manufacture, the top GaAs layer has degraded; apparently due to the formation of a surface oxide or a possible interdiffusion zone.  The sample also has -- the ubiquitous -- surface contamination layer, likely oils and particulates from the atmosphere.  These two extra layers are included in our structural model (in table \ref{table:ranges}), however, for use as a calibration artifact, they are more burdensome then beneficial, the refinement of their parameters are not discussed here.  Only results from the fiducial, unadulterated, buried layers are presented.

\subsection{Parratt formalism}
\label{section:parratt}
XRR modeling has been extensively discussed in literature, including review articles by \citename{chason97} and \citename{russell90}.  The XRR phenomenon arises from subtle changes in electron scattering contrast, or changes in $n_{layer}$, defined as:

\begin{equation}
n_{layer} = 1- r_e\frac{\lambda^2}{2\pi}\frac{\rho_{layer}}{\rho_{bulk}}\sum_a( f_{1a} + if_{2a})N_a
\label{eqn:n}
\end{equation}
Where $r_e$ is classical electron radius, $\lambda$ is wavelength, $f_{1i}$ and $f_{2i}$ is the real and complex components of the atomic scattering factor, $N_a$ is the number density, and $\rho_{layer}$ and $\rho_{bulk}$ are the calculated and bulk mass densities for the layer.  The used \citeasnoun{henke93} scattering factors are available for download (with all recent updates) from \citeasnoun{gullikson12}.

The reflected intensity $I^{R-calc}$, from a film stack, for each $(\theta^i_l, I_{l}^R)$ measurement pair can be modeled as a function of the top layer amplitude reflection ratio, $R_{top}$ (\% coming out), the intensity of X-rays impinging the sample $I_o$ (how many go in), and the instrumental background $I_{bg}$ (how many false or cosmic background counts a detector adds in).  Note that both $I_o$ and $I_{bg}$ are assumed constant over an entire XRR measurement:

\begin{equation}
I^{R-calc}(\theta_l) = I_o |R_{top}(\theta_l)|^2 + I_{bg}
\label{eqn:I}
\end{equation}

We build this $R_{top}$ up from the bottom layers using a series of recursive expressions, which leave us with a calculable, but highly non-linear function. The first and most fundamental part is the Fresnel reflection coefficient, which relates how $n_{layer}$ corresponds to the subtle bending and wavelength changes as X-rays penetrate interfaces between layers. Note this equation assumes each layer has constant $n_{layer}$, (i.e., each layer must have a uniform electron density (homogeneous slab)):

\begin{equation}
r_{layer} = \frac{ k_{\perp layer} - k_{\perp (layer + 1)}}{k_{\perp layer}+ k_{\perp (layer + 1)}}
\label{eqn:Fresnel}
\end{equation}
Where $k_{\perp layer} = 2\pi / \lambda (n^2_{layer} - \cos^2\theta)^{1 / 2}$.

From \citeasnoun{parratt54}, the amplitude ratio for each layer, $R_{layer}$, is derived using the Fresnel coefficients for the current layer $r_{layer}$ and the $R$ term for the next layer down in the stack $R_{(layer + 1)}$.  We have the addition of an oscillation term, $\phi_{(layer+1)}$, for the transmitted wavevector as it propagates through and interferes with, either constructively or destructively, the first reflection (known as the two-beam case).  The phase of the interference is a function of $\theta$ and layer thickness $t_{layer}$:

\begin{equation}
R_{layer} = \frac{r_{layer} + R_{(layer+1)} \phi^2_{(layer+1)}}{1 + r_{layer}\ R_{(layer + 1)}  \phi^2_{(layer + 1)}}
\label{eqn:R}
\end{equation}
Where $\phi^2_{layer} = \exp(i  k_{\perp layer} t_{layer})$.

To complete the recursion after the bottom layer, we assume \emph{a priori} that $R_{substrate} = r_{substrate}$ (i.e., no reflection from the below the substrate interface).  Note that the above equations apply only for perfectly smooth interfaces.  To account for roughness, eqn. \ref{eqn:Fresnel} was modified by \citeasnoun{nevot80} to incorporate an interface width:

\begin{equation}
r_{nevot} = r_{layer} \exp \left [-2(k_{\perp layer} k_{\perp (layer + 1)})^{1/2} Rz_{(layer+1)} \right ]
\label{eqn:nevot}
\end{equation}
Where $Rz_{(layer+1)}$ indicates interface width (i.e., roughness).

For each layer, we have three parameters ($t_{layer}$ from eqn. \ref{eqn:R}), density ($\rho_{layer}$ from eqn. \ref{eqn:n}), and roughness ($Rz_{layer}$ from eqn. \ref{eqn:nevot}).  The substrate only introduces one parameter, $Rz_{substrate}$, as $t_{substrate}$ is assumed to be $\infty$ and the bulk density is assumed to be $\rho_{substrate} = \rho_{bulk} = 5.316$ $g / cm^3$ (for GaAs).  For any given number of layers, $N_{stack}$, we have $d$ fitting parameters, where $d=N_{stack} \times 3 + 1$.

This modeling approach fails dramatically for layers with high roughness or interdiffusion.  Eqn. \ref{eqn:Fresnel} relies on $n_{layer}$ being constant (slab model) which is not the case for layers with interdiffusion.  High roughness, large $Rz$, introduces a high decay rate in eqn. \ref{eqn:nevot} which smooths out interference (thickness) fringes from eqn. \ref{eqn:R} (i.e., oscillations from the $\phi^2$ term).  Alternative modeling approaches are needed for such structures, which is beyond the scope of this study.

\subsection{Data refinement:  optimization vs. sampling}

Optimization methods, such as differential evolution (DE) \cite{storn97} and simulated annealing \cite{solookinejad11}, provide us with fast descent methods to finding a globally good solution to a highly multidimensional (many-parameter) problem.  A random parameter search, such as Monte Carlo sampling, would provide us precision estimates for each parameter in a refinement. However, for industrial applications, the Monte Carlo sampling -- even using efficient methods like MCMC -- is too slow.

\subsection{Optimization cost function}
\label{section:cost}

Our XRR simulations ($I^{R-calc}$, in eqn. \ref{eqn:I}) are fit to the XRR measured data ($I^{R}$), in order to establish the likelihood or fitness (goodness of fit (GOF)) of a given simulation, using a cost function.  We consider the set of all the model's parameters as a $d$ dimensional vector ($\overrightarrow{p}=[p_1, p_2,...,p_d]$).  Fitness is given in terms of $\textbf{p}$, with smaller fitness values indicating better refinement. The lowest fitness from a group $m=\{\textbf{p}\}$ gives the best parameter set, defined as $\textbf{b}$ (for best). 

The least squares method or $\chi^2$ (derived from Gaussian process likelihood function) is commonly used to determine model fitness.  For a detailed explanation of likelihoods and a derivation of $\chi^2$, see section 3.5 of \citeasnoun{sivia06}. $\chi^2$ assumes that measurement errors are independent and uses a Gaussian approximation valid when counts are sufficiently high, $I^R_l > 100$, for every data point in a data set. $\chi^2$ is determined by summing the squared differences between data and model divided by an estimate of the true value for each data point.

\begin{equation}
\chi^2(\overrightarrow{p}) = \frac{1}{N}\sum^{N}_{l=1}\frac{\left[I^R_{l}(\theta_l) - I_l^{R-calc} (\theta_l;\overrightarrow{p})\right]^2}{I_l^{R-calc} (\theta_l;\overrightarrow{p})}
\end{equation}

Note that the denominator should formally be $I^R_l$. However, the introduction of zeros from no (or low) count measurements will cause the fitness to blow up, $I_l^R=0 \implies \chi^2 \rightarrow \infty$, therefore in practice, the denominator in $\chi^2$ is set to $I_{R-calc}$ which always has a finite background, $I_{bg}$.

Although $\chi^2$ is a statistically valid cost function to apply for high count rate data, in the low-count regime  -- common in high angles of XRR measurement -- a log-normal cost function may be more appropriate, as \citeasnoun{sivia06} section 8.6 discusses.  A variant of the log-normal cost function was introduced by \citename{wormington99}, in an early study of alternative cost functions to address the poor sensitivity of $\chi^2$ for XRR analysis. Log-normal mean square error, $MSE_{\log}$, takes the sum of the squares of the differences in the logs of $I^R_l$ and $I^{R-calc}_l$ for each data point:

\begin{equation}
MSE_{\log}(\overrightarrow{p}) = \frac{1}{N-1}\sum^{N}_{l=1}\left[\log I^R_{l} (\theta_l)- \log I_l^{R-calc} (\theta_l;\overrightarrow{p})\right]^2
\end{equation}

Because the $MSE_{\log}$ cost function is found in most commercial XRR packages and is sensitive to high-angle information, it is an excellent choice for intercomparisons. We use this cost function exclusively in this study, $\textbf{p}\equiv MSE_{\log}(\overrightarrow{p})$.

\subsection{Differential evolution}

The DE used here follows the algorithm developed by \citeasnoun{storn97}, first applied to XRR by \citeasnoun{wormington99}, and found today in many commercial and open-source XRR refinement packages; making DE an ideal candidate refinement method for XRR intercomparisons.  Both \citename{storn97} and \citename{wormington99} provide a detailed treatment of the algorithm. Here we use the notation of \citename{wormington99} unless otherwise noted, and provide the  aspects necessary in understanding our strategy in testing the DE technique.  

DE is initialized by selecting a population, $m$ ($NP$ in \citename{storn97}), of possible solution sets $\overrightarrow{p}$ randomly drawn from the allowed ranges for parameters $p$ for a data fitting model.  For our case, we are using an 8 layer structural model, for a total of $ d= 25 = [3 \times 8 + 1]$ parameters in $\overrightarrow{p}$.  The upper and lower bounds for each parameter, $p_i\in \{p_{i,min}\leq p_i \leq p_{i,max}\}$, are given in Table \ref{table:ranges}, and cover a wide (range 1) and two narrow (2 \& 3) ranges. In all cases we assume a uniform likelihood for our random draw of parameters within these bounds (all values, including endpoints are equally likely).  The $m$ of $\{\overrightarrow{p}\}$ are simulated using eqn.s \ref{eqn:n}-\ref{eqn:nevot}, $I^{R-calc}_l(\overrightarrow{p})$, and evaluated for fitness, using $MSE_{\log}(\overrightarrow{p})$ to determine the fittest or `best-so-far' member, $\textbf{b}$ ($best$ in \citename{storn97}), of the DE population.  This initial population of solutions acts as the progenitors of all future generations of solutions.  Wormington suggested a minimum population, where $m > 10 \times d$.   We used a a value of $m=400 = [16 \times 25]$ to provide consistent results.  Performing an optimization of $m$ versus $d$ is beyond the scope of this work, and may be a future research direction.

The DE is then evolved though a series of generations, $G$, to continue to optimize the entire $m$ towards a global minimum for fitness.  In order to select successive generations, $G$, DE uses two strategies to optimize the fitness of $\textbf{b}^G \rightarrow \textbf{b}^{G+1}$, mutation and crossover.  Mutation is the ``differential" aspect of DE.  \citename{storn97} developed a family of possible algorithms, with strategy tailored to the type of system under study.  For this analysis, we adopted the strategy selected by \citename{wormington99} (or $DE/best/1/bin$ in Storn).  In this approach, the difference between two sets of parameters pulled randomly from the current population, $\overrightarrow{p}^G_a,\overrightarrow{p}^G_b\in m^G $, is added to the most fit $\textbf{b}^G$ parameter set. This difference is multiplied by a mutation constant, $k_m$  ($F$ in \citename{storn97}) where $k_m\in [0,2]$; this constant controls the rate of convergence for the algorithm.  A large $k_m$ will more effectively sample the possible parameter space, and a smaller $k_m$ will quickly converge.  \citename{wormington99} suggests $k_m = 0.7$ for XRR refinement, and this value was used exclusively for our study.

\begin{equation}
\textbf{b}^{G+1}= \textbf{b}^G + k_m(\overrightarrow{p}^G_a-\overrightarrow{p}^G_b)
\label{eqn:diff}
\end{equation}

The second strategy, crossover, allows for mixing of $p$ between members of a $m$ into the next generation, and corresponds best to meiosis in genetics.  In this approach, every member of $\overrightarrow{p}^G$ is allowed to exchange individual parameters, $p^G_i$, with the corresponding parameters, $b^G_i$, from the fittest member of the current population, $\textbf{b}^G$, (the Genghis Kahn approach).  Each parameter across all of $m^G$ will swap out with corresponding parameters in $\textbf{b}^G$ at a fixed probability, defined as the crossover or recombination constant, $k_r$ ($CR$ in Storn), where the probability $k_r\in [0,1]$.  Large $k_r$ allows a great deal of parameter mixing and small $k_r$ is closest to asexual reproduction (traditional genetic mixing would be $k_{r}(meiosis) \equiv 0.5$).  For this study, we used \citename{wormington99}'s recommended optimization of $k_r=0.3$.  Lowering this parameter limits intermixing and makes finding hidden minima challenging; raising it can make the solutions' evolution unstable.

One additional check on each new $\overrightarrow{p}^{G+1}$ is to verify that any crossover or mutation satisfies the allowed parameter range criterion, i.e. $p^{G+1}_i\in \{p_{i,min}\leq p_i \leq p_{i,max}\}$  If this is invalid, then a new random parameter is selected for $p^{G+1}_i$. The DE method is highly customizable, with $m$, $G$, $k_c$, $k_r$, $p_{i,min}$, and $p_{i,max}$, all being tunable parameters.  Further, \citename{storn97} introduced other strategies such as using a random member of the population, $\overrightarrow{p}^G_c$, rather than the fittest, $\textbf{b}^G$, in equation \ref{eqn:diff},  and allowing intermixing of more than one member during the mutation and crossover stages.  Exploring the tuning parameters of the DE will be the focus of a future study.

\subsection{Statistical treatment of DE}

In order to develop any meaningful statistical sampling results from an optimization method such as DE, we need to: perform many measurements, perform many refinements, and run each refinement for a very long time.   Although this approach is straight forward, this study involves over one year of computer time to answer structural information questions for a single structure. We ran a large number ($\#$) of DEs, $\#_{DEs}=20$ with DE size $\&$ length of $m=400$, $G=10,000$ for each data set collected and using parameters in table \ref{table:ranges}, range 1. For the 1s set, we collected and refined over 30 XRR measurements (over 600 DE refinements).  (In this paper, only the results from the first ten 1s sets are presented). For the 30 s set, data was refined using Table \ref{table:ranges}, range 1 \& 2 (over 240 DE refinements) and $4\times20$ additional runs were performed for the $I_0$ study (over 480 additional DE refinements). For the 20 s set, data was refined using Table \ref{table:ranges}, range 1 \& 3 (over 120 DE refinements). Each DE refinement ran for approximately 8 hrs (per 2 computer cores).  This large statistical sampling of DE results allows exploration of both the stability of DE refinement for a single data set, and between data sets.  In this way we can assign parameter precision estimations for both DE modeling limits and deviations between data sets, giving us a lower bound for the sensitivity of modeling and the impact of varying noise (data quality).


\section{Results}

The pre-standard's layers were deposited using molecular beam epitaxy; the GaAs and AlAs layers have a 1:1 stoichiometry and atomically sharp interfaces.  As a result, the buried layers exhibit low roughness, seen in minimal decay in interference (thickness) fringes, and densities very close to bulk values.  Table \ref{table:ranges}, range 1, gives the widest parameter ranges in our study.  Layers $Al$ \textit{1, 2,} $\&$ \textit{3} and $Ga$ \textit{1, 2,} $\&$ \textit{3} were all refined using $Rz \in\{0.3 \leq Rz \leq 0.5\}$ nm, corresponding to atomically smooth interfaces.  We allow wide parameter ranges for $t$, $Rz$, and $\rho$, for both the surface contamination layer, $surf$, and the surface-oxidation layer, $Ox$.  We also assert that this oxide layer has a Ga$_1$As$_1$O$_1$ composition and a uniform density, $\rho_{Ox}$, to satisfy eqn. \ref{eqn:Fresnel}.  

The $I_o$ and $I_{bg}$ used in our study were held constant for all XRR measurements in each of the instrument configurations, with the 1 s set at $[I_o=3.5\times10^7$ $\&$ $I_{bg}=2$],  the 30 s set at $[I_o=3\times10^8$ $\&$ $I_{bg}=20]$, and the 20 s set at $[I_o=7\times10^8$ $\&$ $I_{bg}=20]$, respectively (unless otherwise stated).

It may seem counterintuitive, but we need to answer the question of refinement time (how long do I refine my data?) before answering questions of data quality (how long do I count?) and parameter precision estimation (is our result meaningful?).  

\subsection {Refinement lifetime}

The answer for refinement duration, for all complicated models (i.e., large $d$ $\equiv$  ($d>10$)) is to run a large $\#$ of DEs ($\#_{DEs}\geq10$), and run them over long evolution timescales ($G\geq10,000$).  For our $d=25$ DE study, we first attempted $m=200$ and $G=5,000$ for table \ref{table:ranges}, range 1. With $\#_{DEs}= 200$ per data set, and we found only a very small percentage of the results were usable in any context (in most cases not even a single successful DE result per data set ($<1\%$ success rate overall)).  We increased the population and generations ($m=400$ and $G=10,000$) and reproduced the study  for the results presented here.

\subsubsection{Result fitness.}
To answer `how long' we must first address `how fit' our DE must be, in order to be considered a success, (i.e., to have found the model global fitness minima).  
We aggregated the $\textbf{b}_{10,000}$ from each DE set (1 s, 30 s, or 20 s), and found the fittest member,  $\textbf{best}_{10,000} = min\in \{\textbf{b}_{10,000}\}$. \
We then took $\textbf{best}_{10,000}$ and multiplied by $1.1$ to establish a subset range of DEs between the $\textbf{best}$ and ones of almost perfect success  ($1.1\times (\textbf{best}) \equiv \{\textbf{best}_{10,000} \leq \textbf{b} \leq 1.1 \times \textbf{best}_{10,000}\}$), and we repeated using $1.5$ to select a range between the $\textbf{best}$ and `fairly good' success ($1.5\times (\textbf{best}) \equiv  \{\textbf{best}_{10,000} \leq \textbf{b} \leq 1.5 \times \textbf{best}_{10,000}\}$); thus establishing two success measures for our work (see table \ref{table:gof}, columns 1--5 \& 6--10, respectively). 
Note that a new $\textbf{best}_{10,000}$ must be determined for each set of measurements, as the statistical noise present in each (measurement) set may be different. 
 
For the 1 s set, we had over 200 DEs using table \ref{table:ranges}, range 1. If we keep only DEs which meet our $1.1\times (\textbf{best}) $ criterion, we are left with only 14 of the initial 200 DEs ($7\%$, seen in table \ref{table:gof}, column 1).  Likewise, the $1.5\times$ ($\textbf{best}$) criterion results in only $27\%$ of the DEs selected as successful (column 6).  These subsets are later used in parameter precision estimations.  The $1.1\times$ ($\textbf{best}_{10,000}$) and $1.5\times$ ($\textbf{best}_{10,000}$) selection criteria were applied for each DE at evolution times $G=$ 500, 1000, 2,000, 5,000, and 10,000 to determine DE descent (improvement) rate, and to better answer, how long is long enough.  The surprising result for the 1 s set, is that below $G=5000$, none of the DEs meet the more stringent $1.1\times$ (\textbf{best}) selection criterion, and below $G=2,000$ all fail to meet our less stringent $1.5\times$ ($\textbf{best}$) case, (i.e., no successes out of 200 DEs, each DE running for about two hours)! This analysis was repeated for the 30 s (column 2 \& 7) and 20 s (column 4 \& 9) sets. We saw an even lower success rate for the 20 s, low noise data, with both the $1.1\times$  ($\textbf{best}$) and $1.5\times$  ($\textbf{best}$) selection criterion.  In both the 1 s and the 20 s sets (rotating anode measurements), we need a large number of DEs ($\#_{DEs}\geq20$) and long evolution times ($G\geq10,000$) to guarantee finding a global minima for $\textbf{p}$.  For the 30 s sets, we saw a higher percentage of successful DEs, however, we are not sure if this is statistically significant given the still low $\#$ of cases presented.  

\subsubsection{Parameter ranges.}

The difficulty with multidimensional problems in general, and XRR refinement in particular, is the large number of equally possible solutions (many local minima), which all exhibit `good' fitness.  This large number of local minima is the product of using wide allowed parameter ranges and the potential for exchange of $Rz$ and $t$ values between layers in a model. Differential evolution is a successful refinement approach to this type of multimodal problem, as it simultaneously looks in many local minima at once, in order to find the global minima. In the limit where parameter ranges are all significantly narrowed close to the global minima (fittest solution), i.e., $p_i\in\{b_i-\epsilon \leq p_i \leq b_i+\epsilon\}$ where $\epsilon\rightarrow0$, the parameter space becomes unimodal (there is a unique global minimum).  In this limit, the DE should achieve a global best fit within only a few generations (and could even be replaced with more traditional non-linear least squares methods, such as the Levenberg-Marquardt algorithm for solution optimization).  However, finding sufficiently narrow ranges for a $d=25$ parameter model is nearly impossible.  Range 2 (in table \ref{table:ranges}) selects a narrow range of allowed $p_{i, min}$  and $p_{i,max}$ values centered around the $\textbf{best}_{10,000}$ determined with the 30 s set DEs using the wide, range 1 results.  In a similar manner, range 3 selects narrow limits centered around the range 1 $\textbf{best}_{10,000}$ for 20 s set DEs.  In table \ref{table:gof}, we see a significant increase in success rates for both $1.1\times$ ($\textbf{best}$)  and $1.5\times$ ($\textbf{best}$) criteria, columns 3 \& 5, and 8 \& 10 respectively, when we use narrow parameter ranges 2 \& 3 (versus range 1).It is clear from table \ref{table:gof} that for the $1.5\times$ (\textbf{best}) criterion results we achieve a 100$\%$ DE success rate in less than $G=500$ for both the 30 s (column 8) and 20 s (column 10) sets using ranges (2 $\&$ 3).  However, reaching the $1.1\times$ (\textbf{best}) criterion requires $G>2,000$ for the 20 s set and is \emph{never} reached for the 30 s set.  The 30 s result is most vexing, as one of the six XRR measurements in the set is shown to never meet our $1.1\times$ criterion ($83\%$ corresponds to $100\%$ success for 5 out of 6 data sets).

From these results, we arrive at a second conclusion: refinement time can be decreased substantially by narrowing $p_{i, min}$  and $p_{i,max}$ for a given model. Thus, \emph{a priori} structural knowledge (i.e., transmission electron microscopy cross sections) can be used to speed up fitting.  However, setting too-narrow parameter ranges may cause you to miss the global minima entirely.

\subsection {Data collection strategy}

\begin{figure}
    \centering
    \includegraphics[width=.7 \textwidth]{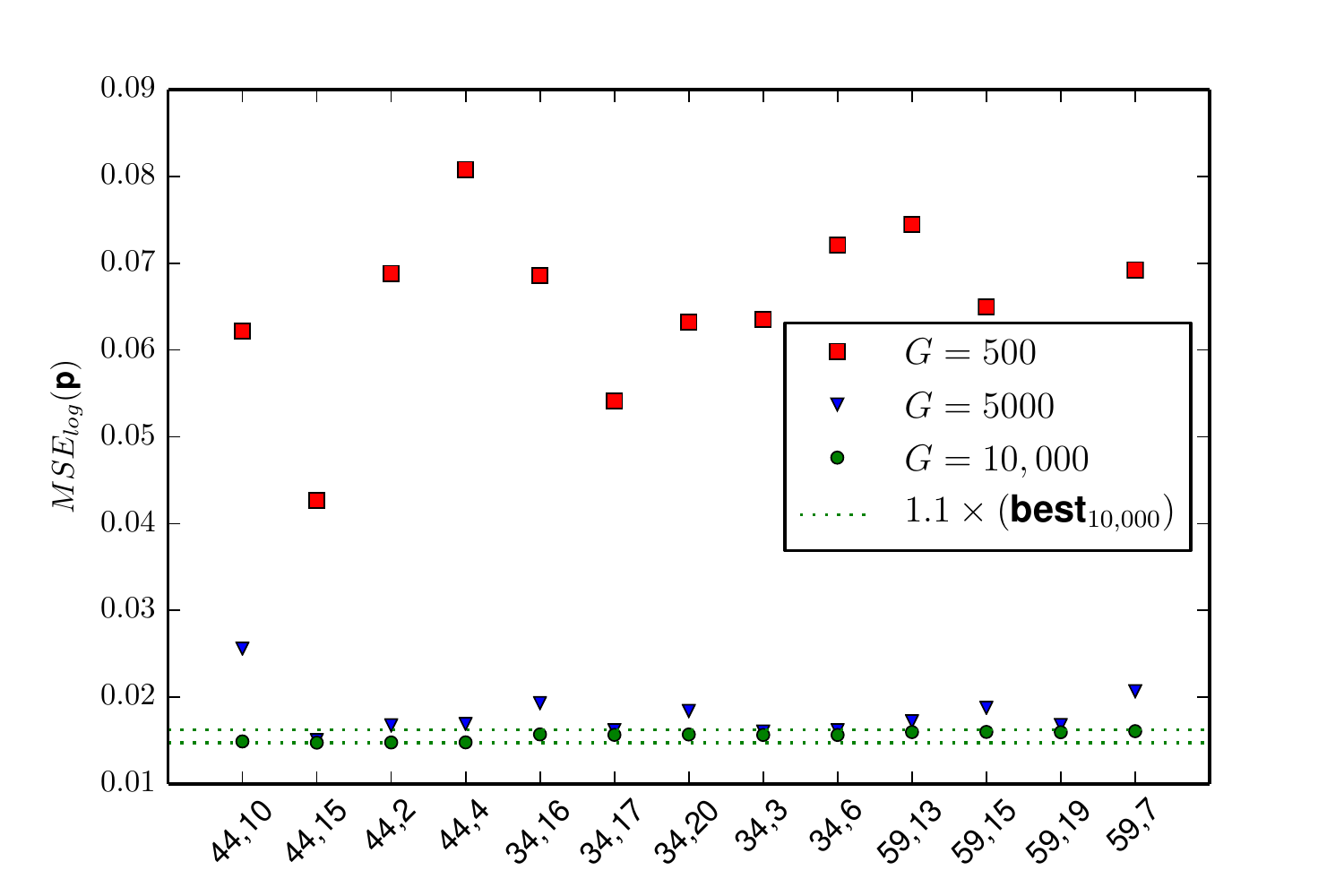}\\
    \includegraphics[width=.7 \textwidth]{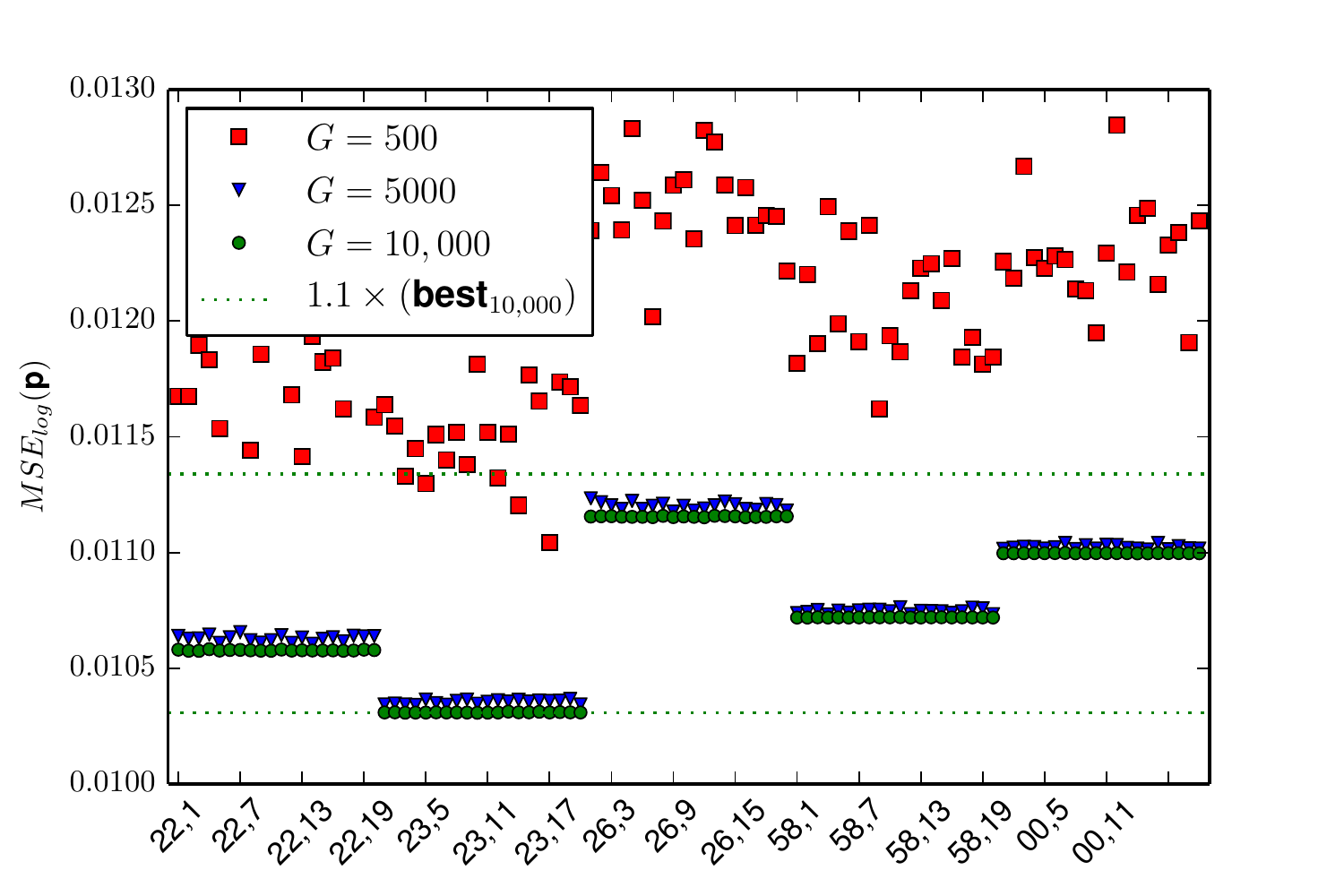}\\
    \caption{Fitness ($MSE_{\log}(\textbf{b}_G)$) for $G=$ 500, 5000, and 10,000 on the 1 s set, wide, range 1 (top) and 30 s set, narrow, range 3 (bottom). These data represent successful solutions meeting a $1.1\times$ $\textbf{best}_{10,000}$ criterion, out of 200, 120, and 120 DEs, respectively.  Note that the 20 s data is not shown, only having 3 successes for this criterion.}
    \label{fig:gof_gens}
\end{figure}

In fig. \ref{fig:gof_gens}, we see the evolution of DE fitness for $G=$ 500, 5,000, and 10,000 on the 1s set, range 1 (top), and on the 30 s set, range 2 (bottom).  The \# of results presented is the \#  of successful DEs out of 200 and 120, respectively. For all of the DEs, $\textbf{b}$ for $G=500$ is higher than for $G=$ 10,000 ($\textbf{b}_{500}\geq\textbf{b}_{10,000}$) indicating that DEs are indeed improving over time, as expected for a descent (to fitness) algorithm. However, there is still a high variance in fitness even for $G=5,000$, $\sigma ^2 (\textbf{b}_{5,000}) \neq 0$. If we define a new term, \textbf{best}$_{data}$, meaning the \textbf{best} DEs from a single measurement, then we clearly see in fig. \ref{fig:gof_gens} (bottom) that \textbf{best}$\not\equiv$\textbf{best}$_{data}$ as there are different minima for each data set. Further, we can answer our $83\%$ success riddle from table \ref{table:gof} (column 3) as the \textbf{best}$_{data\{43\}}$ is higher in fitness than the $1.1 \times (\textbf{best})$ cutoff, i.e., measurement-to-measurement fitness differences are the same order as our aggressive selection criterion, $\sigma ^2 (\textbf{best}_{data}) \cong 0.1 \times (\textbf{best}$).  One way to address data set exclusion when performing precision estimations, is to find $\textbf{best}_{data}$ for each measurement, and apply the criterion for DEs from only that measurement, $1.1\times (\textbf{best}_{data}$).  In this way, \textbf{b} from all measurements will survive within our statistical precision estimations.  $\sigma^2(\textbf{best}_{data})$ is the result of noise fluctuations between measurements (i.e., cosmic events, stray \& scattered photons, and low counting statistics) at large $\theta$ and illustrates the emphasis that $MSE_{\log}(\textbf{p})$ places on low count information. 

\subsubsection {Repeated measurement approach.}

\begin{figure}
    \centering
    \includegraphics[width=.7 \textwidth]{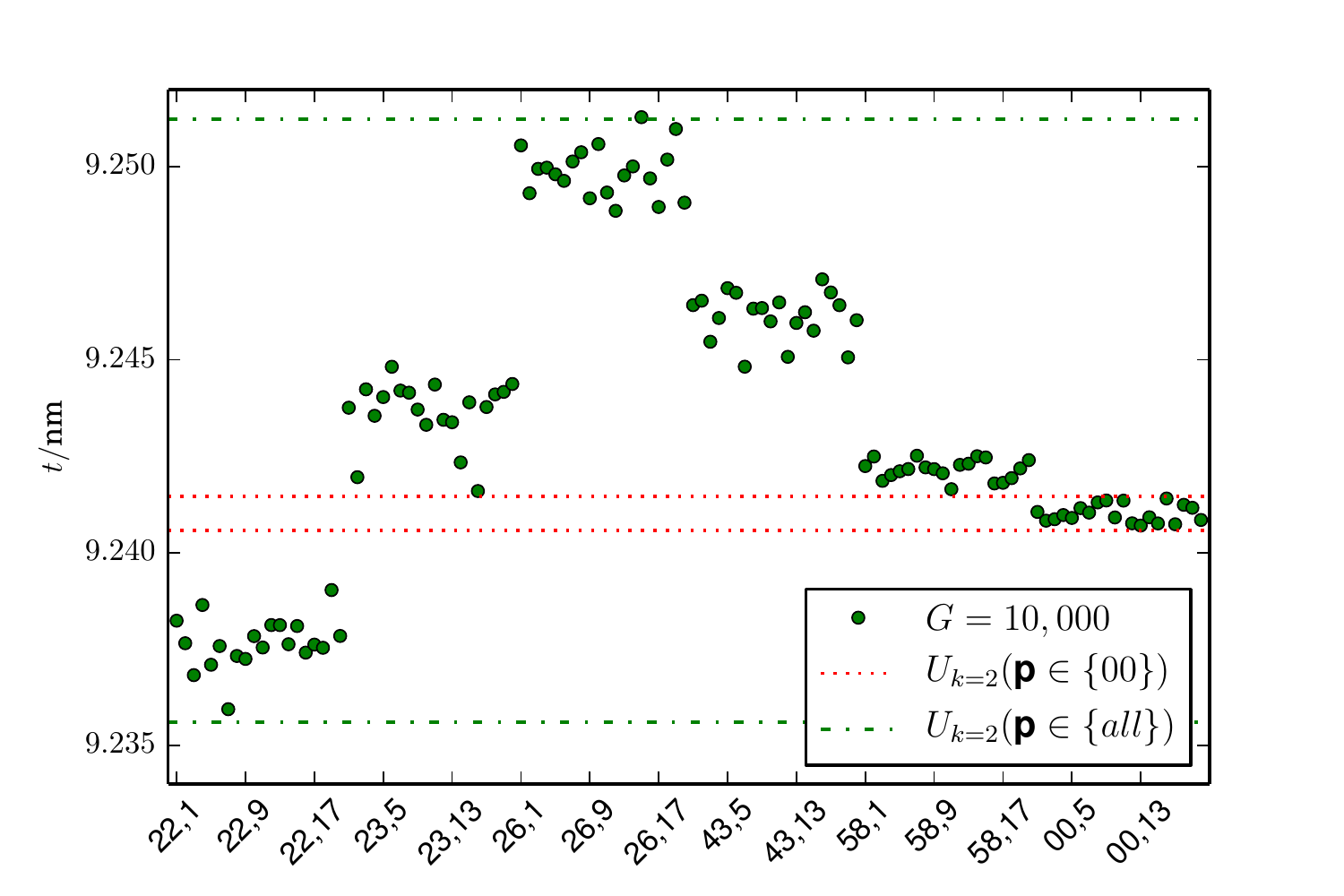}\\
    \includegraphics[width=.7 \textwidth]{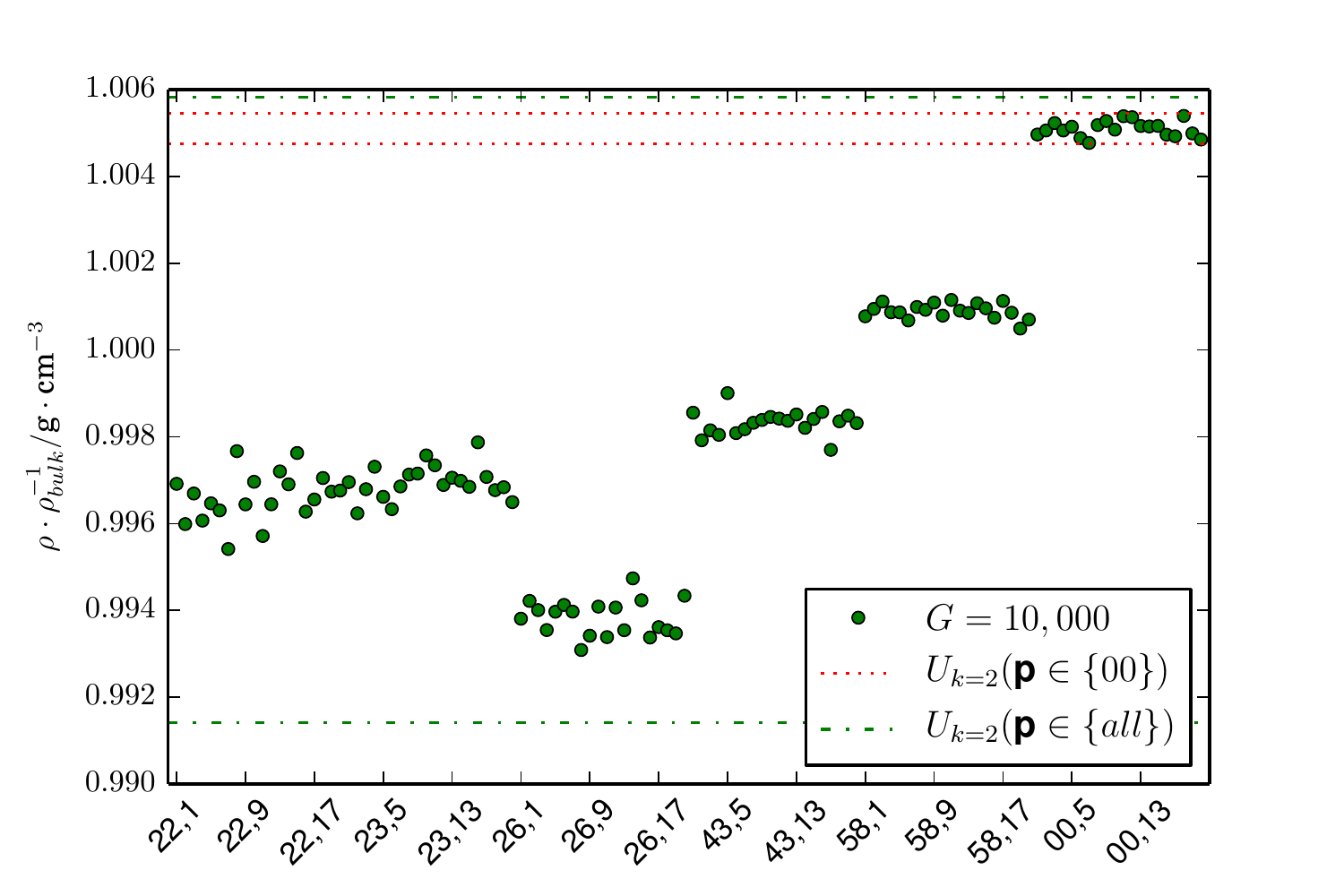}\\
    \caption{The effect of using multiple measurements to establish technique sensitivity for Ga3 (bottom GaAs layer) on thickness (top) and density (bottom).  Results for 30 s set, narrow, range 3 for $1.5\times$ $\textbf{best}_{10,000}$.  Dots are individual DE results for each of the 6 data sets. Dotted, $\cdot \cdot$, lines delineate the precision range for a set of DEs from a single measurement, $\{00\}$. The dash/dot, $- \cdot$, lines are the precision range from all six measurements.}
    \label {fig:GaAs3_00}
\end{figure}

The clear impact of counting statistics noise on fitness suggests our first recommendation for a collection strategy:  Run multiple measurements and run DEs on each measurement.  In this way, we can evaluate sensitivity to measurement noise for each parameter within the model.  Here we will use a $k$ = 2 coverage factor uncertainty estimating method developed in the Guide to the expression of uncertainty in measurement (GUM) to provide XRR modeling precision estimates and measurement-to-measurement repeatability estimates.\cite{gum2008}  In fig. \ref{fig:GaAs3_00}, we again examine the 30 s set, range 2 results, but focus now on on model parameters $t$ \& $\rho$ for \textit{Ga 3} (see table \ref{table:ranges}).  The dotted, $\cdot\cdot$, range shows $U(p_{00})\equiv2\times\sigma(p_{00})$ for the DEs of measurement $\{00\}$.  The dash-dot, $-\cdot$ gives $U(p)\equiv2\times\sigma(p)$ for all DE results.  We clearly see over an order of magnitude increase in parameter refinement precision estimates resulting from noise differences between measurements for both $t$ and $\rho$. Also, the bias between $<p_{00}>$  and $<p>$ is different for $t$ \& $\rho$ (i.e., $<\rho_{00}>$ is near the upper edge of $U(\rho)$, whereas $<t_{00}>$ is closer to the $<t>$).  By examining the $\sigma(p_{data})$, we see that for range 2, $G=$ 10,000 is a sufficiently long timescale for the DEs to approach the global minima ($\sigma(p_{00})\rightarrow 0)$.  So, although we are reaching a global minimum, there exists a bias in $<p_{data}>$ related to data noise, and necessitating multiple measurements ($\#\sim$ 5 to 10) separate this noise induced bias.  

\subsubsection {XRR measurement time.}

In table \ref{table:u_thickness}, we present $U(t)$ (noise-induced precision) for layers \textit{Ga 1}-\textit{Al 3} (from table \ref{table:ranges}) of each of the different measurement sets and normalize these relative to the 30 s data $\{00\}$, range 2 results.  Column 1 provides $<t_{00}>$ and column 2 gives $U(t_{00})$, for DEs of data $\{00\}$; this estimation should correspond only to the residual divergence in DEs from the global minima of $\{00\}$.  Column 3 gives $U(t)$ for the entire 30 s set, range 1 representing measurement-to-measurement noise-induced precision estimates for six, 11 hour, $I_o=3\times10^8$, measurements.  In column 4 we calculate the ratio of this measurement-to-measurement noise induced uncertainty scaled to the refinement uncertainties calculated earlier (column 2). If we average this ratio over all layers (last row) we find a mean($<>$) data noise bias that is $19\times$ greater than refinement precision (column 2).  We perform this analysis again for the 1 s set, range 1, which consists of ten, 22 minute, $I_o=3.5\times10^7$ measurements (columns 5 \& 6) and we find an $<>$ measurement-to-measurement noise to be $23\times$ refinement precision.  Putting this in context, 66 hours versus 3.7 hours of data collection or $18\times$ ($5 \times$ when comparing total $I_o$) yields only a marginal improvement in parameter precision reduction.  If we calculate the same statistics for the 20 s set, which consists of three, 8 hour, $I_o= 7\times10^8$ measurements (columns 7 \& 8) we see an $<>$ measurement-to-measurement uncertainty of $11\times$ versus refinement precision.  In context, this factor of 2 improvement in $U(t)$ required $6 \times$ more data collection time, and still may be artificially low due to the low number of measurements represented.  Table \ref{table:u_density} provides an identical analysis for $\rho$ over the same data sets.  We see a remarkable agreement with the ratios (columns 4, 6, \& 8) for both $t$ and $\rho$.  $U(p)$ results from the 30 s and 1 s sets averaging $\sim20\times$  larger and the 30 s $\{00\}$ results in all cases (with the 20 s averaging $\sim10\times$ larger).  

From this result, it is clear that we can estimate modeling and measurement-to-measurement noise precision using a large set of short scans, such at the 1 s set presented; about 4 hours of rotating anode data.  Higher counting statistics $6\times I_o$ from a sealed tube instrument showed no appreciable benefit, and $20\times I_o$ from the same rotating anode showed only marginal improvement in the noise induced measurement bias.

\subsubsection{Parameter precision estimation.}

\begin{figure}
    \centering
    \includegraphics[width=.7 \textwidth]{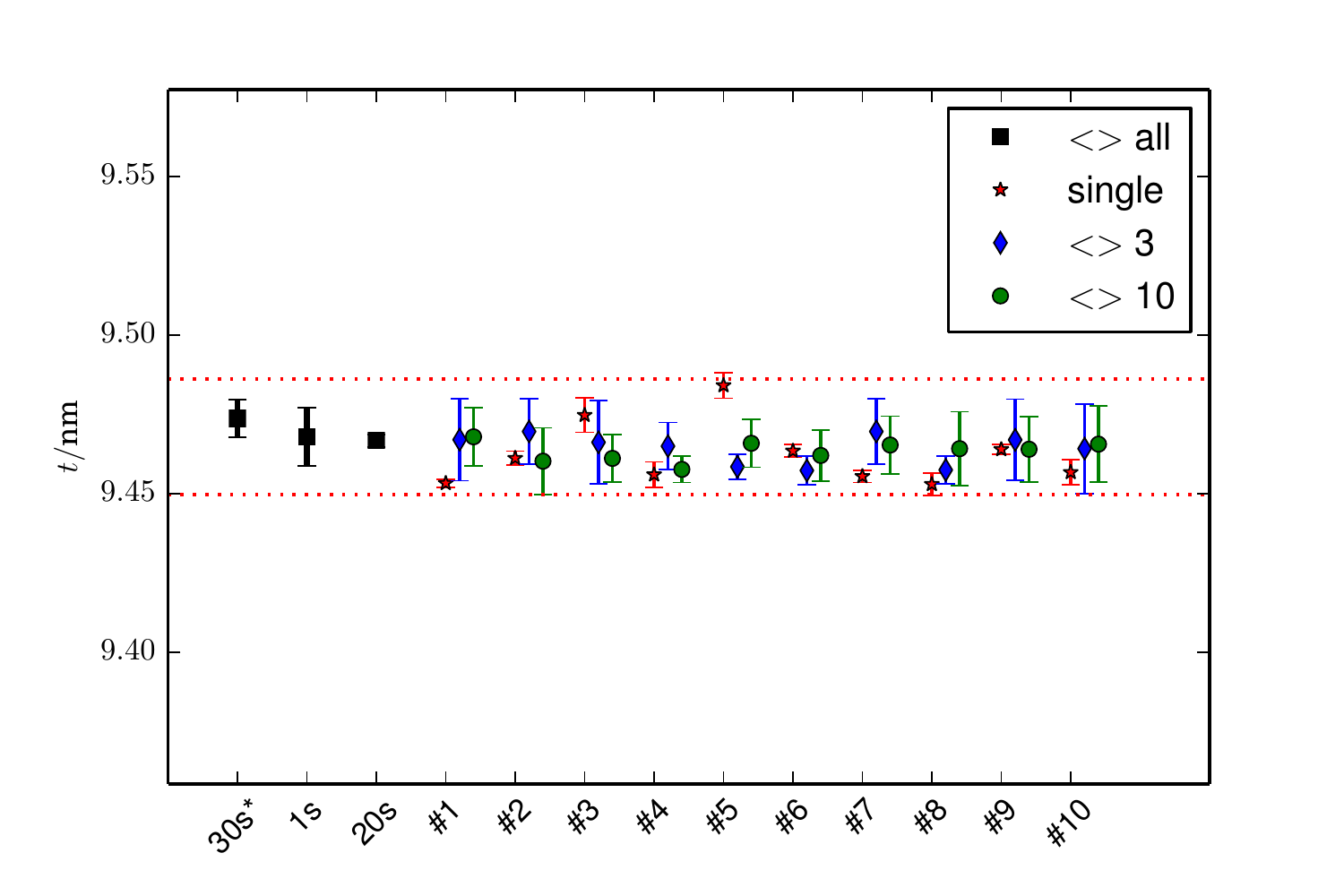}
    \includegraphics[width=.7 \textwidth]{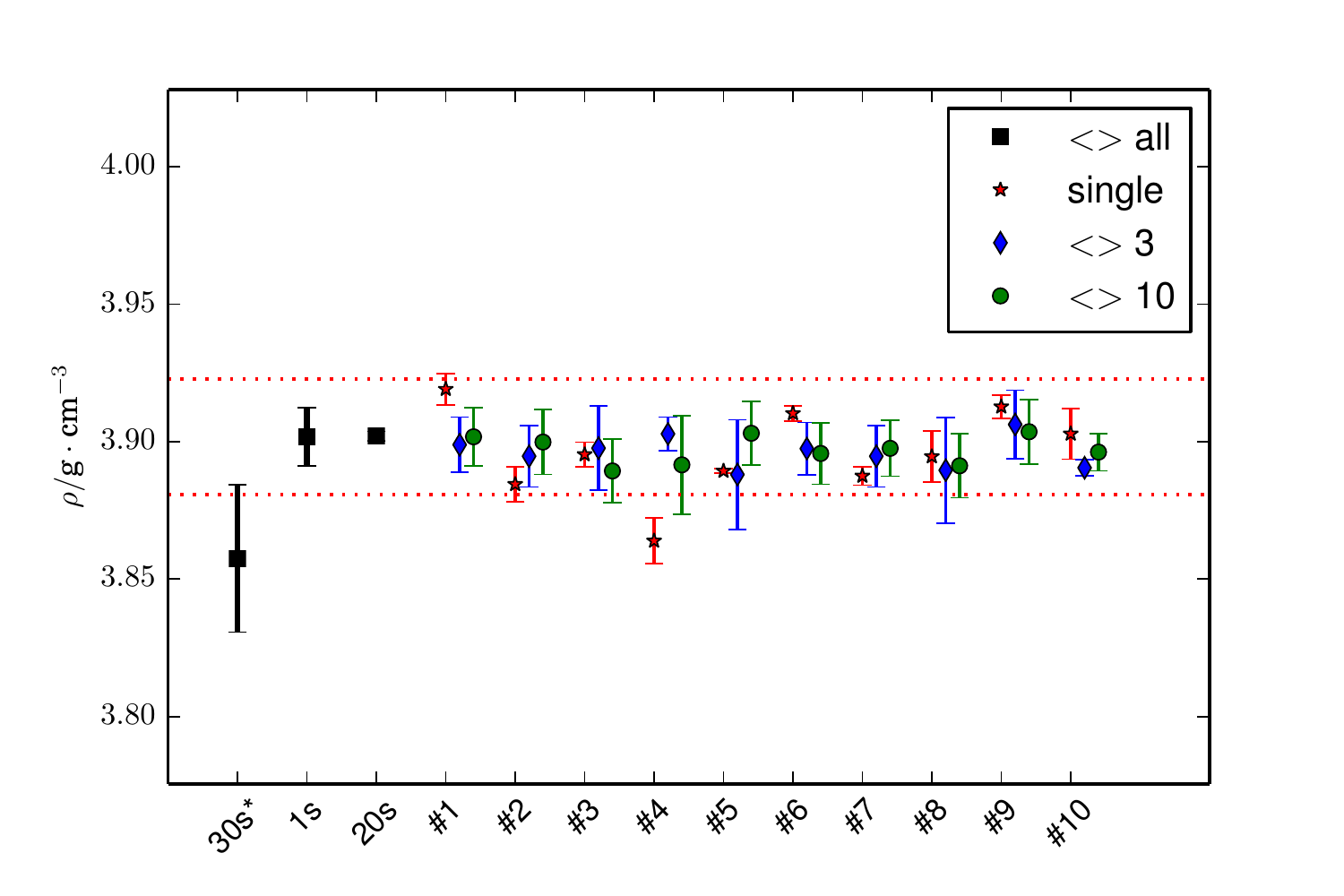}
    \caption{Bottom GaAs layer results for thickness (top) and density (bottom) for 30 s data (from sealed tube instrument), 1 s and 20 s data (from rotating anode) and random draws of 3 and 10 data sets from the set of ten, 1 s XRR data (second data square), with * indicating an alternative method of calculating mean and standard deviation.}
    \label {fig:boostrap_AlAs3}
\end{figure}

Using the 1 s set, range 1, we explored the applicability of developing a measurement-to-measurement precision estimator for XRR.  Our goal here is to find the minimum number of measurements required to produce stable $U(p)$ estimations (i.e., which will effectively reduce noise induced bias from the parameter estimates).   In fig. \ref{fig:boostrap_AlAs3}, we show our parameter estimations for Ga3 $t$ (top) and $\rho$ (bottom), using a random measurement draw (bootstrap) method.  The stars, $\star$, represents the $<p_{data}>$ for each XRR measurement in the set with precision bars providing $u(p_{data})\equiv \sigma(p_{data})$.  Note that these are larger refinement precision than the 30 s, range 2, $\{00\}$ results we saw in tables \ref{table:u_thickness} \& \ref{table:u_density}.  The diamonds, $\diamond$, represent three random data sets drawn from the ten 1 s data sets.  We draw at random 10 times and compare results for $<p>$ and $u(p)$.  We repeated this process for each of the ten sets shown as circles, $\circ$.   The dotted, $\cdot\cdot$, range represents $U(p)= 2\times\sigma(p)$ for the 1 s set (second data square). In comparing the bias from single measurements for $\rho$ (bottom), we see that single data sets may occur outside of $U(p)$ estimates.  It is only after averaging results from 3 measurements that we consistently stay within the bounds, but there is still a high degree of bias.  After comparing the results for all $p_i$, we suggest ten measurements to provide uniformly consistent results for all of the parameters under study.

\subsection {Parameter estimations between instrumentation}

We now have enough tools to try a first look at instrument-to-instrument reproducibility. In tables \ref{table:thickness} \& \ref{table:density}, column 1 are the $<p>$ from our sealed tube instrument and Column 5 \& 6 provide the $\Delta p$ from the $<p>$ from our 1 s and 20 s rotating anode results.  By examining these $\Delta$ we can see that $<t>$ for the rotating anode instrument is nearly constantly outside the $U(t)$ for the sealed tube instrument.  Likewise, $<\rho>$ is nearly always lower for the rotating anode, than for the sealed tube.  These observations suggest that the instrument profile function and instrument alignment systematic bias are often greater than measurement-to-measurement refinement precision, and much greater than refinement precision for single data set.

\section{Conclusions}

Through our analysis of this complex, $d=25$, modeling test case, we are able to develop a list of suggestions for modeling with XRR when using optimization methods to estimate parameter precision for maximizing precision for each tool and reproducibility between tools.

\subsection {Summary of DE $p$ statistical method}
\begin{itemize}
    \item Measure using an instrument with $I_o / I_{bg} > 1\times10^7$
    \item Run a series of XRR measurements $(\# \sim 10)$
    \item Refine DEs with large $m$ ($> 15\times d$) \& long $G$ ($>5000$)
    \item Keep allowed parameter ranges as narrow as possible
    \item Run a large $\#$ of DEs ($\#_{DEs}\geq10$), 
    \item Sort nearly perfect DEs to find $<p>$ \& $U(p)$
     \item Randomly draw XRR results to test for $<p>$ bias
\end{itemize}

If one follows these suggested practices, it is possible to achieve a stable model with precision estimates for parameters in even a complex XRR structure.  This data can then be used to compare reproducibility between XRR instruments.

\subsection{Future directions}

In this work, we neglect the impact of an instrument profile function, sample misalignment, beam footprint, variability in $I_o$ during a measurement, and nonlinearity in detector response, inter alia.
We are currently incorporating instrumental corrections into our modeling software to test these effects for various instrument configurations. 
Our next step will be to use these precision estimates developed here to establish a protocol for using repeated measurements to interrogate instrument-induced bias from instrument-to-instrument configurations through inter-comparisons with the end goal of estimating a combined uncertainty budget. 



\ack{Research performed in part at the NIST Center for Nanoscale Science and Technology.  A special thanks to Kerry Siebein and Chad Snyder for assistance in data collection on instruments at NIST.\\}

\referencelist[JAC_XRR_windover]



\begin{table}
\setlength{\tabcolsep}{5pt}
\caption{GA parameter ranges, $p$, for the DE refinements .}
\label{table:ranges}
\begin{tabular}{lccccc}

LAYER
&COMP.
&PARAM.
&RANGE 1
&RANGE 2
&RANGE 3\\
\hline
 \multirow{3}{*}{\textit{surf}} & \multirow{3}{*}{C} 
& $t / \textrm{nm}$ & 0.5-2.0&1.3-1.7&1.2-1.6\\
&& $Rz / \textrm{nm}$ & 0.1-1.0&0.1-0.4&0.1-0.4 \\
&& $\rho / \textrm{g cm}^{-3}$ & 1.0-5.0&0.5-1.5&0.25-1.75\\ \hline

 \multirow{3}{*}{\textit{O$_x$}} & \multirow{3}{*}{GaAsO} 
& $t / \textrm{nm}$ & 1.0-5.0 & 3.0-3.3 & 3.0-3.2\\
&& $Rz / \textrm{nm}$ & 0.1-1.0 & 0.7-1.0 & 0.8-1.1 \\
&& $\rho / \textrm{g cm}^{-3}$ & 1.0-5.0 & 4.0-4.5 & 3.5-4.5\\ \hline

\multirow{3}{*}{\textit{Ga 1}}&\multirow{3}{*}{GaAs} 
& $t / \textrm{nm}$ & 6.0-11.0 & 8.0-8.5 & 8.3-8.5  \\
&& $Rz / \textrm{nm}$ & 0.3-0.5 & 0.3-0.4 & 0.3-0.4 \\
&& $\rho / \textrm{g cm}^{-3}$ & 2.66-7.98 & 5.3-5.9 & 5.05-5.59\\ \hline

\multirow{3}{*}{\textit{Al 1}}&\multirow{3}{*}{AlAs} 
& $t / \textrm{nm}$ & 9.0-10.0 & 9.4-9.6 & 9.4-9.6 \\
&& $Rz / \textrm{nm}$ & 0.3-0.5 & 0.3-0.4 & 0.3-0.4 \\
&& $\rho / \textrm{g cm}^{-3}$ & 1.91-5.72 &  3.7-4.0 & 3.62-4.0 \\ \hline

\multirow{3}{*}{\textit{Ga 2}}&\multirow{3}{*}{GaAs} 
& $t / \textrm{nm}$ & 9.0-10.0 & 9.2-9.4 & 9.2-9.4 \\
&& $Rz / \textrm{nm}$ & 0.3-0.5 & 0.3-0.4 & 0.3-0.4 \\
&& $\rho / \textrm{g cm}^{-3}$ & 2.66-7.98 & 5.3-5.9 & 5.05-5.9 \\ \hline

\multirow{3}{*}{\textit{Al 2}}&\multirow{3}{*}{AlAs} 
& $t / \textrm{nm}$ & 9.0-10.0 & 9.4-9.6 & 9.4-9.6 \\
&& $Rz / \textrm{nm}$ & 0.3-0.5 & 0.3-0.4 & 0.3-0.4 \\
&& $\rho / \textrm{g cm}^{-3}$ & 1.91-5.72 & 3.6-4.0 & 3.62-4.0 \\ \hline

\multirow{3}{*}{\textit{Ga 3}}&\multirow{3}{*}{GaAs} 
& $t / \textrm{nm}$ & 9.0-10.0 & 9.2-9.4 & 9.2-9.4 \\
&& $Rz / \textrm{nm}$ & 0.3-0.5 & 0.3-0.4  & 0.3-0.4  \\
&& $\rho / \textrm{g cm}^{-3}$ & 2.66-7.98 & 5.2-5.6 & 5.05-5.59 \\ \hline

\multirow{3}{*}{\textit{Al 3}}&\multirow{3}{*}{AlAs} 
& $t / \textrm{nm}$ & 9.0-10.0 & 9.4-9.6 & 9.4-9.6 \\
&& $Rz / \textrm{nm}$ & 0.3-0.5 & 0.3-0.4 & 0.3-0.4 \\
&& $\rho / \textrm{g cm}^{-3}$ & 1.91-5.72 & 3.6-4.0 & 3.62-4.0 \\ \hline

\multirow{3}{*}{\textit{subst}}&\multirow{3}{*}{GaAs} 
& $t / \textrm{nm}$ & -- & -- & -- \\
&& $Rz / \textrm{nm}$ & 0.3-0.5 & 0.35-0.45 & 0.3-0.45 \\
&& $\rho / \textrm{g cm}^{-3}$ & 5.316 & 5.316 & 5.316 \\

\end{tabular}
\end{table}

\begin{table}
\setlength{\tabcolsep}{5pt}
\caption{Percentage of GA optimizations achieving values close to the global minima in goodness of fit (GOF).}
\label{table:gof}
\begin{tabular}{lcccccccccc}

& \multicolumn{5}{c}{\textnormal{||  1.1x BEST GOF  ||}}&
 \multicolumn{5}{c}{\textnormal{||  1.5x BEST GOF  ||}}\\
SET 
&1s
& 30s
& 30s
& 20s
& 20s
& 1s
& 30s
& 30s
& 20s
& 20s\\
RANGE&
 1&
 1&
 2&
 1&
 3&
 1&
 1&
 2&
 1&
 3\\
GEN. $\downarrow$&
(\%)&
(\%)&
(\%)&
(\%)&
(\%)&
(\%)&
(\%)&
(\%)&
(\%)&
(\%)\\
\hline
500&  0& 0&  4& 0& 0& 0& 0& 100& 0& 100\\

1000& 0& 0&  31& 0& 26& 0& 0& 100& 0& 100\\

2000&0& 0& 67& 0& 91& 1& 33& 100& 0& 100\\

5000& 2& 9& 83& 2& 100& 21& 43& 100& 11& 100\\

10000& 7& 46& 83& 3/5& 100& 27& 64& 100& 21& 100\\

\multicolumn{11}{l}{\footnotesize{We present two levels of refinement quality: 1.1x best  GOF which, in some }}\\
\multicolumn{11}{l}{\footnotesize{cases, excludes noise contributions when comparing data sets, and 1.5x}}\\
\multicolumn{11}{l}{\footnotesize{which indicates high quality refinements.  Note that the narrow }}\\
\multicolumn{11}{l}{\footnotesize{best GOF parameter ranges (2 \& 3) all refinements reach 1.5x GOF in the}}\\
\multicolumn{11}{l}{\footnotesize{first 500 generations.}}\\

\end{tabular}
\end{table}


\begin{table}
\setlength{\tabcolsep}{3pt}
\caption{Relative precision in thickness for various collection strategies.}
\label{table:u_thickness}
\begin{tabular}{lcccccccc}

LAYER
& \multicolumn{2}{c}{\textnormal{| 30 s (00)  |}}
& \multicolumn{2}{c}{\textnormal{| 30 s set |}}
& \multicolumn{2}{c}{\textnormal{| 1 s set |}}
& \multicolumn{2}{c}{\textnormal{| 20 s set |}}\\
& $\frac{<t>}{\textrm{nm}}$
&$U(t_{00})$
&$U(t_{30s})$
&$\frac{U(t_{30s})}{U(t_{00})}$
& $U(t_{1s})$
&$\frac{U(t_{1s})}{U(t_{00})}$
& $U(t_{20s})$
&$\frac{U(t_{20s})}{U(t_{00})}$\\

\hline
\textit{Ga 1}&8.298 & 0.001 & 0.014 & 18 & 0.011 & 14 & 0.004 & 5 \\
\textit{Al 1}&9.496 & 0.001 & 0.016 & 21 & 0.024 & 32 & 0.01 & 14 \\
\textit{Ga 2}&9.27 & 0.001 & 0.009 & 15 & 0.009 & 14 & 0.011 & 17 \\
\textit{Al 2}&9.486 & 0.001 & 0.011 & 21 & 0.015 & 30 & 0.005 & 10 \\
\textit{Ga 3}&9.241 & 0.0 & 0.01 & 22 & 0.013 & 30 & 0.008 & 19 \\
\textit{Al 3}&9.462 & 0.001 & 0.012 & 14 & 0.019 & 23 & 0.004 & 5 \\
$<>\downarrow$&--& -- & -- & 19 & -- & 23 & -- & 11\\

\end{tabular}
\end{table}


\begin{table}
\setlength{\tabcolsep}{3pt}
\caption{Relative precision in density for various collection strategies.}
\label{table:u_density}
\begin{tabular}{lcccccccc}

LAYER
& \multicolumn{2}{c}{\textnormal{| 30 s $\{00\}$  |}}
& \multicolumn{2}{c}{\textnormal{| 30 s set |}}
& \multicolumn{2}{c}{\textnormal{| 1 s set |}}
& \multicolumn{2}{c}{\textnormal{| 20 s set |}}\\
& $\frac{<\rho>}{ \textrm{g}\cdot \textrm{cm}^{-3} }$ 
&$U(\rho_{00})$
& $U(\rho_{30s})$
&$\frac{U(\rho_{30s})}{U(\rho_{00})}$
&$U(\rho_{1s})$
&$\frac{U(\rho_{1s})}{U(\rho_{00})}$
& $U(\rho_{20s})$
&$\frac{U(\rho_{20s})}{U(\rho_{00})}$\\

\hline
\textit{Ga 1}&5.442 & 0.002 & 0.055 & 22 & 0.058 & 23 & 0.016 & 7 \\
\textit{Al 1}&3.897 & 0.003 & 0.07 & 23 & 0.034 & 11 & 0.02 & 7 \\
\textit{Ga 2}&5.323 & 0.002 & 0.051 & 27 & 0.049 & 25 & 0.028 & 15 \\
\textit{Al 2}&3.878 & 0.002 & 0.049 & 25 & 0.027 & 14 & 0.011 & 5 \\
\textit{Ga 3}&5.347 & 0.002 & 0.033 & 18 & 0.039 & 21 & 0.015 & 8 \\
\textit{Al 3}&3.905 & 0.002 & 0.054 & 27 & 0.031 & 15 & 0.003 & 2 \\
$<>\downarrow$&--& -- & -- & 23 & -- & 18 & -- & 7\\

\end{tabular}
\end{table}


\begin{table}
\setlength{\tabcolsep}{5pt}
\caption{Differences in thickness for various collection strategies.}
\label{table:thickness}
\begin{tabular}{lcccc}

LAYER
& \multicolumn{1}{c}{\textnormal{30 s}}
& \multicolumn{1}{c}{\textnormal{30 s \{00\} }}
& \multicolumn{1}{c}{\textnormal{ 1 s }}
& \multicolumn{1}{c}{\textnormal{ 20 s }}\\
& t  / \textrm{nm}
&$\Delta t$
&$\Delta t$
&$\Delta t$ \\

\hline

\textit{Ga 1}&8.308 &  -0.01 & 0.04 $\uparrow$ & 0.028 $\uparrow$ \\
\textit{Al 1}&9.496 &  0.0 & 0.022 $\uparrow$& 0.026 $\uparrow$\\
\textit{Ga 2}&9.271 &  -0.001 & 0.016 $\uparrow$ & 0.007 \\
\textit{Al 2}&9.495 &  -0.009 & -0.007 & -0.0 \\
\textit{Ga 3}&9.241 &  -0.0 & 0.014 $\uparrow$ & 0.017 $\uparrow$\\
\textit{Al 3}&9.474 & -0.011 & -0.012 & -0.007 \\

\end{tabular}
\end{table}


\begin{table}
\setlength{\tabcolsep}{5pt}
\caption{Differences in density for various collection strategies.}
\label{table:density}
\begin{tabular}{lcccccc}

LAYER
& \multicolumn{1}{c}{\textnormal{30 s}}
& \multicolumn{1}{c}{\textnormal{30 s \{00\} }}
& \multicolumn{1}{c}{\textnormal{ 1 s }}
& \multicolumn{1}{c}{\textnormal{ 20 s }}\\
& $\frac{\rho}{\textrm{g cm}^{-3}}$
&$\Delta \rho$
&$\Delta \rho$
&$\Delta \rho$ \\

\hline

\textit{Ga 1}&5.678 &  -0.236  $\downarrow$ & -0.201 $\downarrow$ & -0.192  $\downarrow$ \\
\textit{Al 1}&3.922 &  -0.025 & -0.094$\downarrow$ & -0.103 $\downarrow$\\
\textit{Ga 2}&5.557 &  -0.234 $\downarrow$ & -0.162 $\downarrow$ & -0.168 $\downarrow$ \\
\textit{Al 2}&3.815 & 0.063 $\uparrow$ & -0.034 & -0.023 \\
\textit{Ga 3}&5.391 &  -0.043 $\downarrow$ & -0.073 $\downarrow$ & -0.078 $\downarrow$\\
\textit{Al 3}&3.858 & 0.048 & 0.038 & 0.044 \\

\end{tabular}
\end{table}

\begin{table}
     \setlength{\tabcolsep}{5pt}
     \caption{Definitions for common terms.}
     \label{table:def}
     \begin{tabular}{ll}
$I_o$ & Incident X-ray intensity (fixed)\\
$I_r$  & Reflected intensity (measured or calculated)\\
$I_{bg}$ & Background intensity of instrument (detector noise)\\
$d$ &  \# of structural model parameters, $p$ in the set, $\overrightarrow{p}$ \\
$n_{layer}$ & index of refraction of a layer\\
$t_{layer}$ & Thickness of a layer\\
$\rho_{layer}$ & Density of a layer\\
$Rz_{layer}$ & Roughness of a layer\\
$p$ & A single parameter from the structural model\\
$\overrightarrow{p}$ & All parameters from the structural model\\
$\textbf{p}$ & Fitness of $\overrightarrow{p}$ using the cost function\\
$m$ &  The population of $\textbf{p}$ in each DE $G$\\
$G$ &  Last generation of the DE (counting from G=0)\\
DE &  Algorithm optimizing $\textbf{p}$ fitness of $m$ over $G$\\
$k_m$ & Mutation constant or convergence rate for DE\\
$k_r$ & Recombination rate (parameter exchange rate) for DE\\
$\textbf{b}_G$ &  The fittest member of $\textbf{p}$ from $m$ for $G$\\
$\textbf{best}$ &  The fittest member of a set of DEs for XRR data sets\\
$\textbf{best}_{data}$ & The fittest member of a set of DEs for one data set\\
$1.5 \times (\textbf{best})$ &  Subset of almost perfect $\textbf{b}$ from a DE result set \\
$1.1 \times (\textbf{best})$ &  Subset of fairly good $\textbf{b}$ from a DE result set \\
$<p_i>$ & Mean value for $\{p_i$\}\\
$\sigma(p_i)$ & Standard deviation of $\{p_i\}$ (also $u(p_i))$\\
$U(p_i)$ & Precision estimate using $k$ = 2 coverage ($\equiv 2\sigma(p_i)$)\\
\{\} & A set of either $p$ or XRR measurements\\
     \end{tabular}
\end{table}




\begin{figure}
    \centering
    \includegraphics[width=.7 \textwidth]{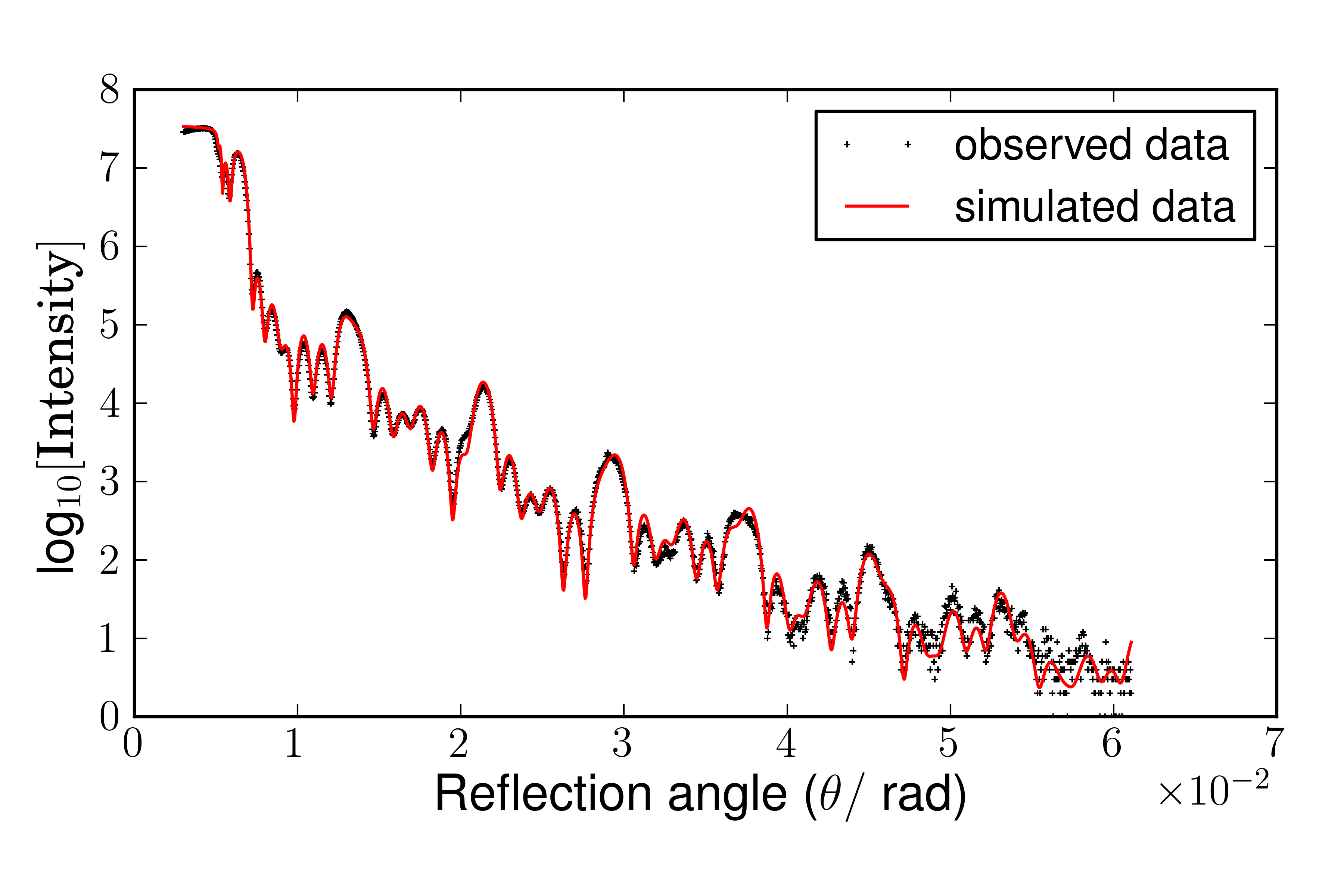}\\
    \includegraphics[width=.7 \textwidth]{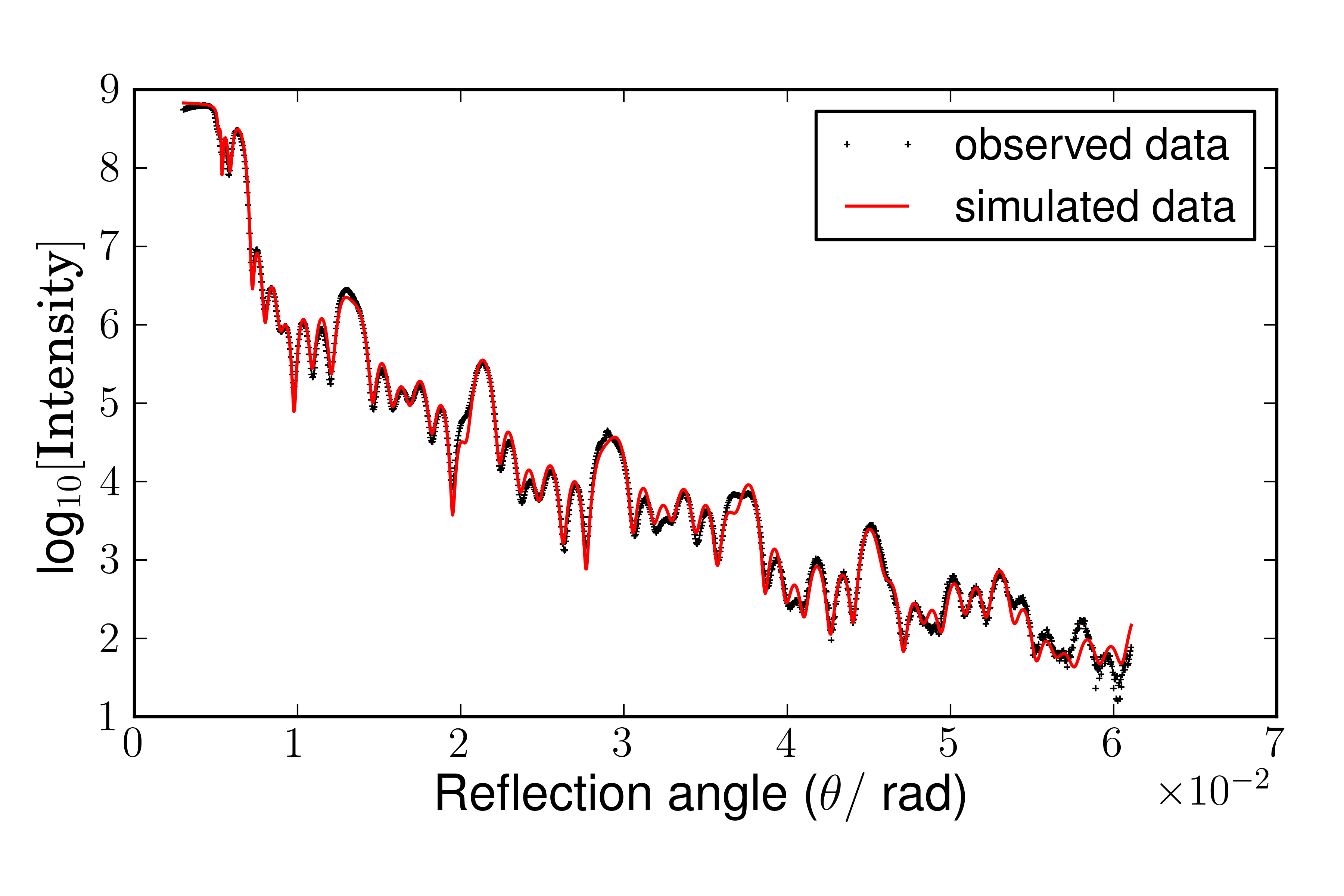}\\
    \caption{XRR measurement (crosses) and best fit model (line) for 1s data (top) and 20 s data (bottom), both showing successful refinements.}
    \label{fig:XRR}
\end{figure}

\end{document}